\documentstyle[11pt,cal97,psfig]{article}

\begin{document}

\title{Flat-fielding of ACS WFC grism data}

\author{N. Pirzkal, A. Pasquali and J. Walsh}
\affil{ST-ECF, Karl-Schwarschild st-2, Garching bei Muenchen, D-85748, Germany}



\begin{abstract}
In direct imaging, broad band flat-fields can easily be applied
to correct deviation from uniform sensitivity across the detector field. However
for slitless spectroscopy data the flat field is both field and
wavelength dependent. The effect of the wavelength dependent flat
field for the slitless G800L grism mode of the ACS Wide Field Channel
(WFC) has been investigated from observations of a flux calibrator at
different positions in the field. The results of various flat-fielding
schemes are presented including application of a flat field cube
derived from in-orbit broad band filter flat fields. Excellent results
are reported with deviations in the extracted spectra $<2\%$ across 
the WFC field.
\end{abstract}




\section{Available WFC flat-fields}

New ACS in-orbit filter flat-fields, were recently constructed using globular cluster observations (Mack et al., 2002).  These flat-fields show that there is a significant amount of large scale structure which varies from one broad band filter to another (i.e with wavelength). Flat-fielding ACS G800L grism data is therefore required in order to be able to derive a unique G800L sensitivity curve which is the same at all positions on the detector. We have used these flat-fields and have fitted them using a 3rd order polynomial as a function of wavelength and at each pixel. The result of this fit was the creation of a data cube enabling aXe, the ST-ECF Slitless Extraction Software (Pirzkal et al., 2001), to compute the proper flat-fielding coefficient at any pixel position and wavelength.  Unfortunately, the in-orbit LP flats do not offer an ideal sampling of the wavelength dependence of the flat-field because: 1) they are broad; 2) only a limited number of filters are available; and, 3) there is no filter which reaches wavelengths beyond 8500\AA~ (while the G800L grism mode remains sensitive to about 10,000\AA). 

\section{Observations}

Two white dwarfs have been observed using the WFC G800L grism mode: GD153 during the ACS SMOV campaign and G191B2B during the Cycle 11 Interim Calibration. GD153 was observed at five different positions which were read-out using ACS subarrays. Unfortunately, the target position did not always end up at the center of the subarray and this resulted in only two useful positions where the first order spectra were completely within the subarray. The spectra at the other positions were truncated and did not contain all the flux from GD153 at each wavelength.  G191B2B, was observed using the same technique but at 9 different positions, 5 in the top part of the detector (CHIP1) and 4 in the bottom part (CHIP2). The G191B2B spectra  were all completely within each aperture and each position was observed twice for a total of 18 spectra across the detector. Figure 1 shows the positions and names associated with each of the G191B2B observations.

\begin{figure}
\hspace{1.5in}
\psfig{file=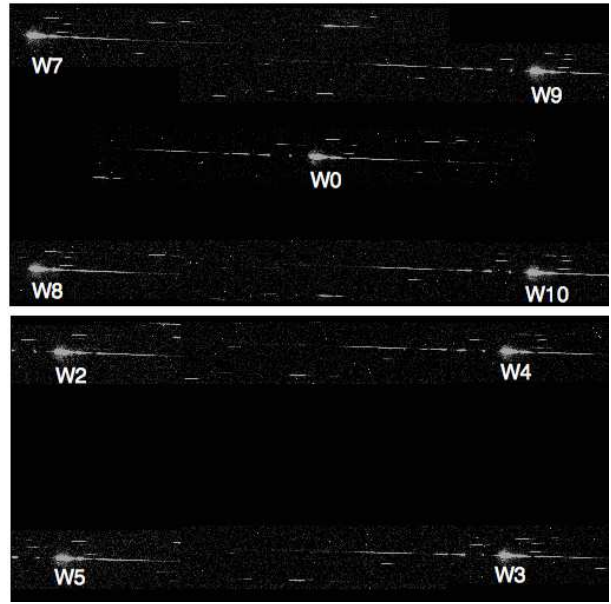,width=8cm}
\label{fig1}
\caption{Mosaic of the nine G800L subarray observations of the White Dwarf G191B2B. The content of this image was bias and dark subtracted and gain corrected. The first order spectrum is immediately above the labels in this figure. The second and third orders are to the right of the first order, while the zeroth and negative orders are to the left of the labels.}
\end{figure}

\begin{figure}
\hspace{1.5in}
\psfig{file=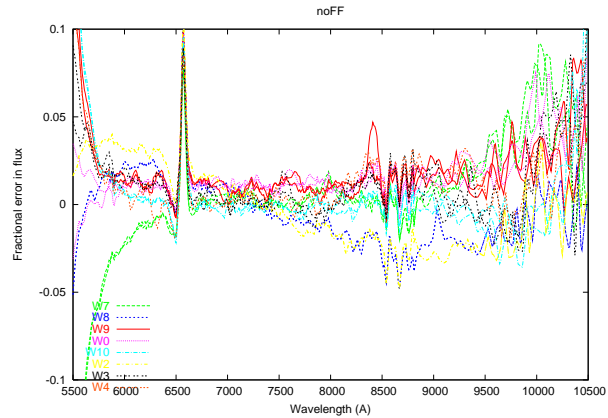,width=8cm,angle=-90}
\label{fig-2}
\caption{The fractional error in the extracted, calibrated, but {\em not} flat-fielded spectra of G191B2B is shown. In the region where the grism is most sensitive (6000\AA~ to 9200\AA) the inconsistencies in flux levels can be seen to progressively increase from about $\pm1.5\%$ to as much as $\pm 3\%$ with, more importantly, some large ($5\%$) systematic differences between different parts of the detectors. Note that observations taken at the same positions are plotted using the same line type and are at a closely similar flux error, demonstrating the high level of repeatibilty one can expect from the G800L grism mode. The increasing errors below 5500\AA~ and beyond 10000\AA correspond to wavelength ranges where the sensitivity of the grism approaches zero. The peak around 6500\AA~ corresponds to the H$\alpha$ absorption feature of G191B2B. The larger errors at the position of  W7 at wavelengths below 6500\AA~ is caused by a less well known wavelength calibration in that part of the detector}
\end{figure}

\begin{figure}
\hspace{1.5in}
\psfig{file=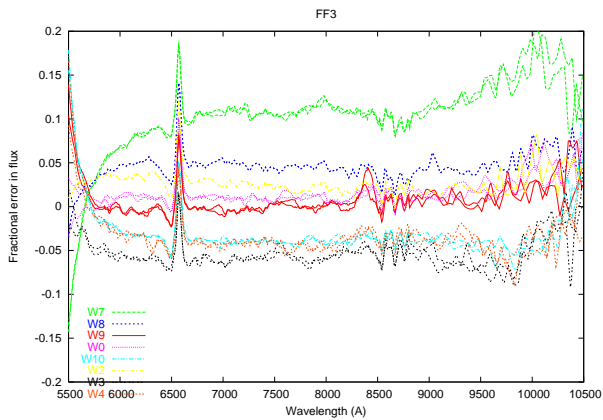,width=8cm,angle=-90}
\label{fig-3}
\caption{The fractional error in the extracted, calibrated, flat-fielded spectra of G191B2B (using a flat-field data cube derived directly from the latest in-orbit flats)., to be compared with Figure 1. The effect of simply applying broad band filter flat-fields clearly introduces some rather large-scale differences across the detectors. These differences are introduced by the direct imaging flats which correct for the field variation of the pixel effective size, which is related to the geometric distortion of the WFC and which affects spectroscopic data differently from direct imaging data.}
\end{figure}

\section{aXe extraction}
\subsection{No flat-fielding}
A first extraction was performed using aXe and no flat-fielding. The extracted spectra of G191B2B were otherwise fully calibrated in physical flux units using the latest estimate of the G800L sensitivity curve.  Figure 2 shows the fractional flux difference between the aXe extracted and calibrated spectra and a template spectrum of G191B2B. The spectra observed at the same positions agree with the template spectrum very closely even though they were obtained at different times. There is however an increasingly large observed discrepancy in the measured fluxes at wavelengths beyond 7500\AA. This effect is both field, and wavelength dependent.

\subsection{Direct LP-flat data cube}
A second extraction of the G191B2B spectra was performed using the data cube fit of the new in-orbit flats mentioned above. The result of which is shown in Figure 3. A significant difference is apparent between different positions on the detector. This are several reasons why this should be expected.  First, it is likely that the large scale flat (L-flat) characteristic of the G800L grism is  different than that of the broad filter flats used to generate the flat-field data cube. A large scale L-flat correction to these broad band filters is hence understandable. Second, the broad band flat-fields are designed so that a direct image of the sky will look flat, even though the effective pixel size of the WFC varies significantly from one corner of the detector to another. Applying such a flat-field introduces a correction which is related to the geometric distortion and which should be corrected for differently in spectroscopic data. The effect of distortion and tilt between the grism assembly and the detector is accounted for by a field dependence of the dispersion properties of the G800L  grism (Pasquali et al. 2002).

\subsection{Corrected LP-flat data cube}
The variation in observed fluxes between different positions on the detector was successfully fitted to a quadratic surface. This relation, essentially an G800L empirical L-flat correction, was used to generate a new, G800L-corrected flat-field data cube. Extracted G191B2B spectra using this new G800L-corrected flat-field curve are shown in Figure 4. Note that this flat-field cube only allows for a wavelength variation of the flat field at wavelengths ranging from 4350\AA~ to 8500\AA. Beyond this range, no gain in accuracy is expected as a constant flat-field coefficient was used. This new corrected flat-field cube produces spectra which reach flux calibration accuracies close to the  $1\%$ level across a wide range of wavelengths, and across most of the detector. The same G800L-corrected flat-field cube was used to extract the two un-truncated SMOV observations of GD153, described earlier, and similar accuracy was achieved between the extracted and flux calibrated spectra and the template spectrum of GD153.

\begin{figure}
\hspace{1.5in}
\psfig{file=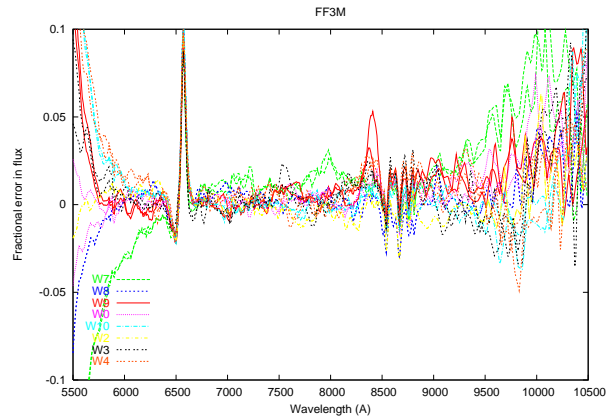,width=8cm,angle=-90}
\label{fig-4}
\caption{The fractional error in the extracted, calibrated, flat-fielded spectra of G191B2B (using our {\em G800L-corrected} in-orbit LP flats-field cube). At wavelengths smaller than 8500 \AA~ where a wavelength dependent flat-fielding was applied, the error in measured flux is within $1\%$ and flat-fielding has removed the systematic differences between different positions on the detector as well as the wavelength dependence visible in Figure 2. Using this flat-fielding scheme, a unique G800L sensitivity curve can be computed and applied to all the extracted spectra.}
\end{figure}

\section{Conclusion}
We have successfully constructed a G800L flat-field data cube which allows one to reach flux calibration accuracies of about $1\%$ at wavelengths ranging from 6000\AA~ to 9000\AA~ and across most of the WFC field of view . This modified G800L-corrected flat-field cube can be used directly by the extraction software aXe to extract un-drizzled, non-geometrically corrected grism observations. It will be made available from the ST-ECF ACS spectroscopy group at http://www.stecf.org/instrument/acs/ . 



%
%

%

\end{document}